\documentstyle[12pt,aasms4,psfig]{article}

\def\etal{{ et~al.\ }}
\def\eg{{ e.~g.\ }}

\righthead{Corrugated Velocity Pattern in NGC 5427}
\normalsize

\begin{document}

\title{Detection of a Corrugated Velocity Pattern in the Spiral
Galaxy NGC 5427.
\footnote{Based on observations made with the William Herschel
Telescope
operated on the island of La Palma by the Isaac Newton Group in the 
Spanish Observatorio del Roque de Los Muchachos of the 
Instituto de Astrof\'\i sica de Canarias.}}

\author{
Emilio J. Alfaro\altaffilmark{2}, Enrique P\'erez\altaffilmark{2},
Rosa M. Gonz\'alez Delgado\altaffilmark{2}, Marco A.
Martos\altaffilmark{3} and Jos\'e Franco\altaffilmark{3}}

\altaffiltext{2}{Instituto de Astrof\'\i sica de Andaluc\'\i a
(CSIC), Apdo. 3004, 18080 Granada, Spain.\\
Electronic mail: emilio@iaa.es, eperez@iaa.es, rosa@iaa.es}

\altaffiltext{3}{Instituto de Astronom\'{\i}a, Universidad Nacional
Aut\'onoma de M\'exico, Apdo. Postal 70-264, 04510 M\'exico D. F.,
M\'exico.\\
Electronic mail: pepe@astroscu.unam.mx, marco@astroscu.unam.mx}

\begin{abstract}
Here we report the detection, in H$\alpha$ emission, of a radial
corrugation in the velocity field of the spiral galaxy NGC 5427. The
central velocity of the H$\alpha$ line displays coherent, wavy-like
variations in the vicinity of the spiral arms. The spectra along
three different arm segments show that the maximum amplitude of the
sinusoidal line variations are displaced some 500 pc from the central
part of the spiral arms. The peak blueshifted velocities appear some
500 pc upstream the arm, whereas the peak
redshifted velocities are located some 500 pc downstream the arm. 
This kinematical behavior is similar to the one expected in a galactic
bore generated by the interaction of a spiral density wave with a thick
gaseous disk, as recently modeled by Martos \& Cox (1998).
\end{abstract}
 
 
\section{Introduction}

It has been known for decades that the gaseous disk of our Galaxy
displays coherent vertical distortions (Kerr 1957; Gum, Kerr \&
Westerhout 1960). In particular, for locations inside the solar circle, 
a wavy structure has been observed in the vertical distribution of
several interstellar tracers, including the atomic and ionized components
(Lockman 1977; Sanders, Solomon \& Scoville 1984; Spicker \& Feitzinger
1986; Malhotra 1995), and young stellar objects (\eg Dixon 1967; Alfaro
\etal 1992a,b; Berdnikov \& Efremov 1993). These undulations, distributed
above and below the mean Galactic plane and in both the azimuthal and
radial directions, have been termed as ``corrugations'', and they seem to
be common features appearing in disk galaxies. They have also been
detected in both the gaseous and stellar disks of other spiral galaxies
such as M31 (Arp  1964), NGC 4244, and NGC 5023 (Florido \etal 1991). 

Their origin could be ascribed to either well localized features, such as
spiral arms and collisions with high-velocity clouds (HVC), or large
scale perturbations, such as gravitational interactions. For instance,
some authors have explored the possibility of undulations along spiral
arms, induced by magneto-gravitational instabilities (\eg Nelson 1985;
G\'omez de Castro \& Pudritz 1992; Kim \etal 1997; Franco \etal 2000). 
In this case, the arms are corrugated by the undular mode (Parker 1966)
of the instability, and the structuring occurs $only$ in the azimuthal
direction (a possible association to the radial patterns could be made if
one assumes that these are the projection of the azimuthal corrugations
on the radial direction). Another localized origin could be ascribed to
collisions of HVC with the disk of a galaxy: HVC impacts could cause the
midplane disk to oscillate in a mode of wavy patterns as well (Franco
\etal 1988; Santill\'an \etal 1999). In the large scale scenario, the
undulations represent the natural response of the disk, which behaves
like a loose drumhead, to the perturbations induced either by some
companion galaxies or other galactic subsystems (Weinberg 1991; Edelsohn
\& Elmegreen 1997).

One obvious inference is that these corrugations should be associated
with equally undulated coherent motions, and models addressing the
vertical displacements also show wave-like $z$-velocity fields with
amplitudes of the order of tens of km~s$^{-1}$ (\eg Nelson 1976,
1985; Franco \etal 1989, 2000; Edelsohn \& Elmegreen 1997; Kim \etal
1997; Santill\'an \etal 2000). Despite this expectation, however, the
data about actual velocity features associated with spatial
corrugations are scarce. The observational evidence for small-scale
distortions in the vertical velocity fields of disk galaxies comes
from the so-called {\sl rolling motions} in our Galaxy (observed in
the $b - V$ plots of the HI galactic distribution; Yuan \& Wallace
1973; Feitzinger \& Spicker 1985), the velocity field of the ionized
gas in M 51 (Goad, De Veny \& Goad 1979), rotation curves of some
spirals (Pismis 1965; Rubin, Ford \& Thonnard 1980), and optical and
radio isovelocity maps of some field galaxies (Bosma 1978; Pismis
1986). In particular, the study of the velocity field of the ionized
gas in M 51 performed by Goad, De Veny \& Goad (1979) suggests that
part of the observed nonrotational velocities could be due to motions
perpendicular to the plane of the disk. Pismis (1965, 1986), on the other
hand, has pointed out the relevance of analyzing the location of the
velocity undulations in the rotation curve with respect to the spiral
structure of the host galaxy. Clearly, all this kinematic information may
be important in our understanding of corrugations.

In this paper we report the detection of a corrugated velocity field
in the ionized gas of the nearly face-on spiral NGC 5427. The central
velocity of the H$\alpha$ emission line shows velocity shifts with
similar behavior when the gas passes through the spiral arms. This
behavior can be understood as due to a hydraulic jump, or bore,
generated by the interaction of the spiral density wave with the gas
of a thick gaseous disk, as discussed by Martos \& Cox (1998,
hereafter MC) and Martos \etal (1999). The detection of this velocity
pattern in NGC 5427 does not answer the current questions about the
morphogenesis of corrugations, but represents a powerful tool to
explore further the theoretical and observational aspects of the
problem. The paper has been organized in three sections, the first
one being this introduction; section 2 is devoted to the description
of the observations and reduction procedure and, finally, in section
3, we discuss the results and outline the main conclusions.

\section{Observations and data reduction}

NGC~5427 is a nearly face-on spiral, interacting with the spiral
NGC~5426. The two galaxies in this binary system have the same
morphological classification, SA(s)c, in the RC2 (de Vaucouleurs, de
Vaucouleurs \& Corwin 1976), and they are cataloged as object 276 in the
atlas of Arp
(1966), and object 21 in the catalogue of Vorontsov-Velyaminov (1959).
The system has been studied photometric and spectroscopically by Blackman
(1982). The inclination of NGC~5427 with respect to the plane of the sky
is about 30$^{\circ}$, and the position angle for the line of nodes is
about 70$^{\circ}$ (Blackman 1982; Grosbol 1985). The star formation
properties of NGC 5427 are studied by Gonz\'alez-Delgado \& P\'erez
(1992, 1997) and Gonz\'alez-Delgado \etal (1997), who have analyzed the
H$\alpha$ emission and the luminosity and size distribution functions.
The nucleus of NGC 5427 is classified as a Seyfert type 2, and Colina
\etal (1997) have studied the relative importance of the circumnuclear
star formation and nuclear activity.

The observations were done with the 4.2m William Herschel Telescope
at the Observatorio del Roque de los Muchachos, during a director's
service night, 18 March 1990. We used the red arm of the ISIS
spectrograph with an EEV CCD chip and the grating R600. A spectral
range of 900 \AA\ was covered with a dispersion of 0.74
\AA~pixel$^{-1}$, centered at 6560 \AA. The sampling along the
spatial direction is 0.33 arcsec~pixel$^{-1}$, equivalent to 56
pc/pixel \footnote{According to the NED IPAC Database, we adopt a
recession velocity of 2618 km s$^{-1}$ for NGC 5427.}, thus for $\rm
H_0=75~km\,s^{-1}\,Mpc^{-1}$, the scale is 170 pc arcsec$^{-1}$.
An exposure of 2000 s was secured through a 4 arcmin long, 1 arcsec
wide slit at a position angle of 45$^{\circ}$ across the nucleus
(Fig. 1).

The data were reduced using the FIGARO data processing software. Bias
and flat-fielding were performed in the usual way. A two dimensional
wavelength calibration was done using the program ARC2D (Wilkins \&
Axon 1991) by fitting a third order polynomial to the position of the
lines in the calibration lamp frames; the rms deviation from the fits
was less than 0.1 \AA. The FWHM of the sky lines [O I] $\lambda
\lambda$6300,6363 measured in the wavelength calibrated frame of NGC
5427 is 1.51$\pm$0.06 \AA\ and 1.57$\pm$0.08 \AA, thus the spectral
resolution is better than 1.6 \AA. The spectrum was flux calibrated
using a flux standard observed through a very wide slit, and the sky
background subtracted.

We used the program LONGSLIT (Wilkins \& Axon 1991) to interactively
perform a gaussian fit to the emission lines along the slit. The
program output includes the position and intensity of the emission
lines, together with the respective fitting errors. The rms error in
velocity due to the wavelength calibration is 1.3 km~s$^{-1}$, smaller
than errors in the gaussian fit. In the subsequent manipulation of
the data, we propagate the errors accordingly. The final errors are shown
in figures 2a,b where the main contribution comes from the gaussian fit
uncertainties.

\section{Results and interpretation}

Figure 2a shows the deprojected radial velocity of NGC~5427,
V$_{deproj}$, together
with the full intensity curves for the emission line H$\alpha$ along
the slit. The velocity curve has been corrected for the galaxy  
systemic motion, 
the projection
with respect to the plane of the sky (30$^{\circ}$) and the P.A. of
the line of nodes (70$^{\circ}$). Our data is more extended towards
the north-east, away from the interaction zone with NGC~5426. The
southern half of NGC~5427 is significantly affected by the presence
of NGC~5426 but, in contrast, the northern half of the galaxy does
not show any significant structural perturbation and seems to be
relatively undisturbed. Thus, the north-south velocity asymmetries are
likely due to the perturbation induced by NGC~5426. The asymmetry in
the star forming properties is also interesting, and it has been
discussed by Blackman (1982) and Gonz\'alez-Delgado \& P\'erez (1992).
The figure shows a remarkably well defined wave-like structure in the
central velocity, superimposed to a gentle rise towards the north-east.
Figure 2b shows the variations in the deprojected velocity $\Delta
V_{deproj}$ (see below), when the slit passes through the 3
spiral features located at galactocentric distances from 5 to 10 kpc
in the north-east. The amplitude of the variations $\Delta$V$_{deproj}$ 
is in the range $\rm\pm20\ to\ \pm30\ km\,s^{-1}$ for
the three main features. In this figure, the velocity has been
de-trended by fitting a linear component to the gentle rise, and is
presented upside-down with respect to Fig. 2a: so that blueshifted
material is seen higher in the velocity axis.

The cross-correlation of the velocity and intensity curves for these
three features shows that the peak velocities are displaced from the
peak intensities by the same distance in all three cases. This is
shown in Figure 3 (full line), together with the auto-correlations
for the velocity (dotted) and intensity (dashed) curves: a negative
lag indicates a shift toward smaller galactocentric distances. The
FWHM of the auto-correlation functions is 560 pc for the intensity,
and 910 pc for the velocity. The velocity versus intensity
cross-correlation function is slightly asymmetrical, with a peak at
300 pc and a full width at zero intensity centroid at 500 pc. Thus
the displacement between the peak velocities and the arm peak
intensities amounts to about a half of the arm width.

An inspection of the WPFC2 image from the HST archive, taken through
the F606W filter (Figure 4) shows the location of the dust lanes
along the spiral arms. In the central regions (inside $r\sim 25$
arcsec), the dust lanes are located in the concave (interior) side of
the spiral arms. For the regions external to 25 arcsec, however, the
position of the dust lanes depends on the interaction with NGC~5426.
For the NE region, which is relatively unperturbed and the residual
velocity curve is better defined, the dust lanes are clearly located
in the convex (exterior) side of the arms. In contrast, the heavily
perturbed SW region displays a different pattern and, even as far as
50 arcsecs away the center, the dust lanes appear in the concave
(interior) side of the arm. These north-south asymmetries, as stated
above, are due to the velocity perturbations induced by NGC~5426 and
the co-rotation radius cannot be defined in a unique form in this
galaxy. For the purposes of the present study, however, we restrict
the analysis to the relatively unperturbed NE region and use the 
location of the dust lanes to infer the direction of gas flow through
the arms.

The wavy structure of $\Delta$V$_{deproj}$ represents the projection of
the residual galactic velocity on the line of sight, once corrected from
systemic and rotational velocities. The system velocity has been   
estimated at the position of the maximum emission at the core.
$\Delta$V$_{deproj}$ can be written
in terms of the inclination angle, $i$, and the angle between the
position vector and the line of nodes, $\phi$, as
$$ \Delta V_{deproj}=\Delta V_{\bot} \,cos(i)+\Delta
V_{\Vert}\,sin(i) , \eqno(1)$$
and assuming that the parallel component  is only due to rotation 
$$ \Delta V_{deproj}=\Delta V_{\bot} \,cos(i)+ \,r\, \Delta \Omega 
\,sin(i)\,cos(\phi) ,\eqno(2)$$
where $r$ is the galactocentric radius, V$_{\bot}$ is the vertical 
velocity component, and $\Omega$ is the angular rotation speed in the
local galactic frame. According to this expression there are three
main factors (or any combination of them) that could be producing the
undulation of the residual velocity: (i) The galactic disk is not
planar and $i$ changes with galactocentric radius. (ii) The galaxy
has a corrugated vertical velocity component. (iii) The rotation
curve of the gas is accelerated and decelerated in a cyclic way. The
inclination of NGC~5427 ($\sim$30$^{\circ}$) is small enough to place
vertical motions nearly on the line of sight and yet large enough to
exhibit the rotational motions. From the second term in eqn. (2), and
even when a good fraction of the velocity corrugation is likely
vertical, one cannot discard the presence of velocity components
parallel to the disk. 
 
Figure 2b shows that the approaching (negative values) peaks in
$\Delta V_{deproj}$ occur in the convex border of the arms, whereas
the receding maxima (positive values) are located behind the
H$\alpha$ emission maxima, in the concave side. An approaching
velocity means that the gas is either flowing up to higher galactic
latitudes, or the rotation velocity is decelerating. On the other
hand, the receding velocities indicate that the gas is either falling
down onto the galactic plane, or is accelerating. Given that the FHWM
of the velocity auto-correlation function is almost twice the FWHM of
the H$\alpha$ intensity auto-correlation, it seems as if the gas flow
is decelerating and jumping above the arm and depositing material in
the concave side. This behavior is similar to the response of a gas
flow into a spiral density wave in a thick and magnetized gaseous
disk, as recently discussed by MC. They found that the gas entering a
spiral arm rises suddenly on the upstream side of the arm, then
accelerates and bends downward, finally landing on a large downfall
region downstream of the arm. Figure 5 illustrates this dynamical
behavior for case 28 of MC, in which the radial extent of the arm is
1 kpc. The center of the horizontal axis, $x=0$, corresponds to the
minimum of the potential well; $z$ is the vertical distance above the
midplane, and the magnetic field is oriented perpendicular to the
$x-z$ plane, along the long axis of the arm. The upstream ascent and
downstream downfall of the gas above the arm are so predicted by the
model in this numerical MHD simulation.

This scenario fits quite well the phenomenology that we find in NGC 5427.
The observed deprojected velocity amplitudes are of the same order as
the ones found by MC. Figure 5 shows that one should expect that the
approaching velocity maxima would appear in the upstream side of the
arms;
for NGC~5427 these maxima are located in the convex borders. Thus, if the
observed velocity pattern is induced by a spiral density wave the gas is
entering the arm through the convex side, what can only occur outward of
the co-rotation radius in a trailing arm. This is supported by the
geometrical arguments described above. Thus, the main features observed
in the radial corrugated velocity pattern seem to be explained by the
perturbations originated in the velocity field pattern of gas
encountering a spiral arm. Such a velocity field occurs on both sides of
the galactic midplane, generating doubled emission lines. The relative
strengths of these two components depend on the relative position of the
gas centroid with respect to the center of the spiral potential minimum,
but the extinction through the disk and the low emissivity of gas in the
interarm regions can make it difficult to detect, or separate, the
emission generated on the far side of the disk. The optical thickness of
spiral galaxies is highly variable across the disk; it is optically thin
in most parts of the interarm regions and much more opaque along the
arms, particularly in the dust patches (e.~g. Disney, Davies \& Phillipps
1989; Marziani \etal 1999). Our spectroscopic data passing through the
three spiral features of the NE region could reach visual extinctions
between 1-4 magnitudes (Marziani \etal 1999). At present we cannot
provide specific extintion values along the arms, but it is possible
that the H$\alpha$ emission generated at the back side is below our
detection limit. The opacity in the interarm regions is certainly much
lower, but the emissivity of the ionized gas is also much smaller and the
signal to noise ratio is accordingly lower. Nonetheless, there is a hint
that we may be detecting some of the far side emission because the FWHM
of the emission lines increases in the high velocity features. Figure 6
shows that the emission with $| \Delta V | > 10$ km s$^{-1}$ has
larger dispersion values, but the low signal to noise ratio prevents us
to discern between foreground and background velocity peaks (we thank the
referee for pointing out this possibility). More work with higher quality
data is needed to conclusively resolve this issue.

Even though the galactic bore model seems to provide a very appealing
explanation to the observed velocity corrugations in NGC~5427, other
mechanisms could also be inducing or enhancing the corrugations of
the velocity field. The fact that this galaxy is a member of an
interacting pair, whose companion has a mass similar or larger than
NGC 5427, makes it likely that the perturbations originated by the
tidal interaction will largely affect the velocity field, producing
corrugated patterns (Edelsohn \& Elmegreen 1997). In addition, the
azimuthal corrugations in the arms originated by magneto-gravitational
instabilities (Franco \etal 2000) must also be present and superimposed
with the other deviations of the velocity field. We notice, however, that
the velocity field resulting from the Parker instability will be
esentially that of a downfall (towards the midplane) along the bent
magnetic field lines, and not the rising and falling motions expected
from the galactic bore picture.

Summarizing, here we report the detection of a  corrugated
velocity pattern in the disk of NGC~5427. A detailed analysis of 
this velocity structure
shows a remarkable good qualitative agreement between the observed
features and those derived from the galactic bore model described by
MC and Martos \etal (1999). Given that the only requirements for this
mechanism to work are the existence of a spiral density wave and a
magnetic field (with a strength similar to those observed in our
Galaxy; see MC and discussion therein), the generation of radial
velocity corrugations must be a common occurrence in spiral galaxies.
The search for vertical velocity corrugations in a sample of face-on
galaxies is now a must follow up program. The difference with the
previous situation is that now we know how and what to search for.  

{\bf Acknowledgments}:
We thank Don Cox for helpful discussions, and Arcadio Poveda for pointing
out the early works of Paris Pismis on undulations of the rotation
curves of spirals. We are indebted to an anonymous referee for a very
useful and detailed report that have helped us to improve the contents 
of the paper. We are grateful to Dave L. King who made the spectroscopic
observations in service mode. This research has made use of the
Starlink Software Collection of the UK PPARC, and of the NASA/IPAC
Extragalactic Database (NED) which is operated by the Jet Propulsion
Laboratory, California Institute of Technology, under contract with
the National Aeronautics and Space Administration. The HST image of
NGC~5427 was retrieved from the Space Telescope European Coordinating
Facility HST archive. 
EJA acknowledges
financial support granted by CONACYT-M\'exico and CSIC-Spain under
the agreement of both institutions; his work is partially supported
by the Spanish DGICYT through grants PB97-1438-C02-02 and by the
Research and Education Council of the Autonomous Government of
Andaluc\'{\i}a (Spain). EP and RMGD are supported by Spanish DGICYT
grant PB98-0521. JF thanks the Instituto de Astrof\'{\i}sica de
Andaluc\'{\i}a for its warm hospitality, and acknowledges partial
support by DGAPA-UNAM grant IN130698 and by a R\&D CRAY Research
grant.

\clearpage

%
%

\clearpage

 \figcaption{Continuum subtracted H$\alpha$ narrow band CCD image of NGC 5427.
North is up and east to the left. The slit used for the spectroscopic 
observations is drawn for reference, at a position angle of 45$^{\circ}$.}

\figcaption{(a) Velocity (full dots with error bars) and intensity
(dotted line) curves obtained from the H$\alpha$ emission line along the
slit. The velocity (left ordinate scale) has been corrected for the disk
inclination (30$^{\circ}$) and the P.A. of the line of nodes
(70$^{\circ}$), and the systemic velocity has been subtracted. The
spatial scale is computed with a systemic velocity of 2618 km s$^{-1}$
and H$_0$=75 km s$^{-1}$ Mpc$^{-1}$. The H$\alpha$ intensity scale is
drawn in the righthand side ordinate axis. (b) An expansion of the two
curves around the zone between 5--10 kpc, where the relationship between
the velocity and intensity structures is clearest; the velocity curve has
been de-trended for the gentle rising slope and plotted upside-down from
(a), with the approaching blueshifted velocities higher in the ordinate
axis.}

\figcaption{Cross-correlation function (full line) of the two curves
shown in Fig. 2b. The velocity (dotted) and intensity (dashed)
auto-correlation functions are also plotted. The two central pixels of
the velocity auto-correlation function are produced by noise and should
not be taken into account; notice then how the velocity auto-correlation
is significantly wider than the intensity.}

\figcaption{Image of NGC~5427 taken with the WPFC2 trough the F606W 
filter. The location of dust lanes is apparent as white patches running
along the spiral arms. The line segment marked in the main North spiral
arm shows the position at which the dust lanes change from the internal
(concave) side to the external (convex) border. Our analysis is done in
the outer regions of the NE part of the spiral structure, where the dust
lane is located in the convex side of the arm.} 

\figcaption{Density contours (15 levels, from 0.01 to 3 times the initial
midplane density) and velocity field for gas approaching a spiral arm
from the left (upstream) side of the numerical grid, case 28 of MC at an
elapsed time of 67 Myr. The size of the longer velocity vector in the
field corresponds to a speed of 55 km s$^{-1}$.}

\figcaption{Line widths (FWHM) versus $| \Delta V |$ for the spectral
features used to determine the velocity curve. Note how for $|\Delta V |
\geq$ 10 km s$^{-1}$ the FHWM of the lines is always larger than 35 km
s$^{-1}$. This widening could be originated by the blending of two
unresolved features comming from both sides of the galaxy.}   








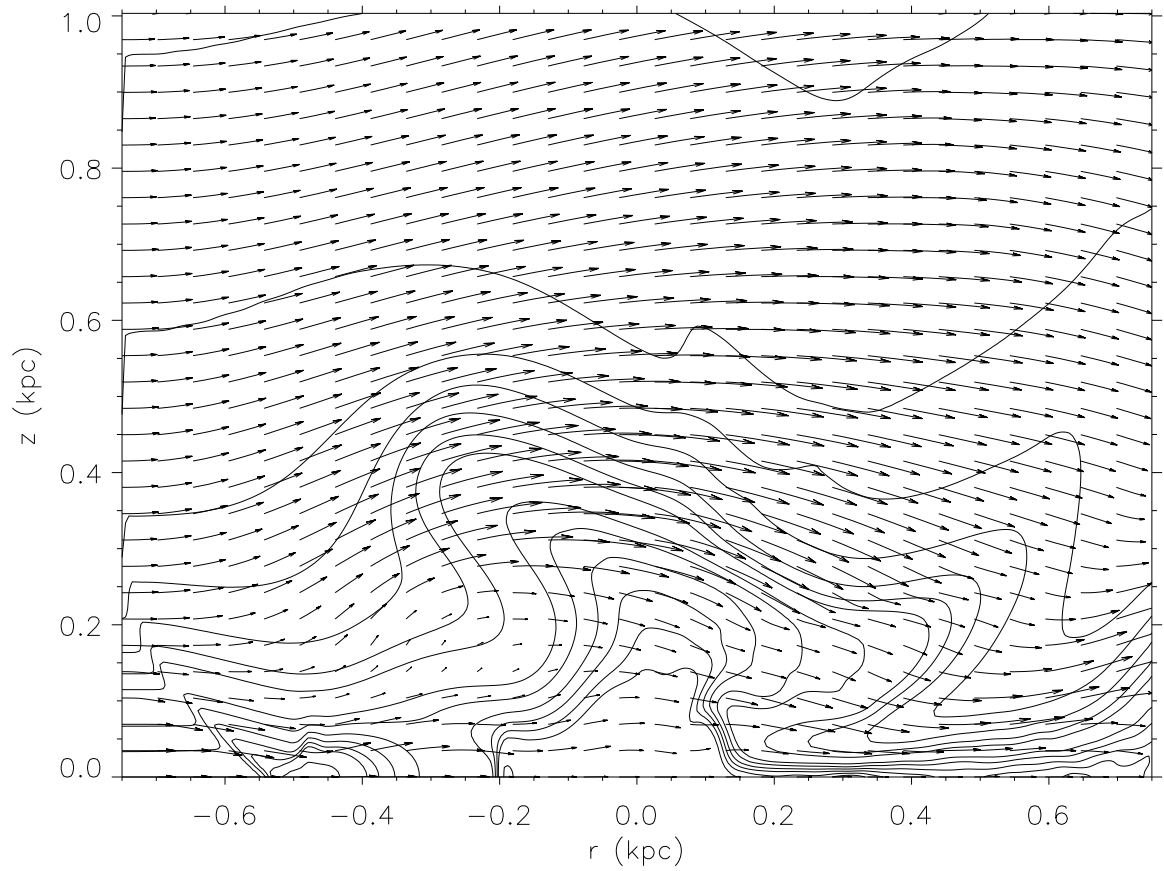
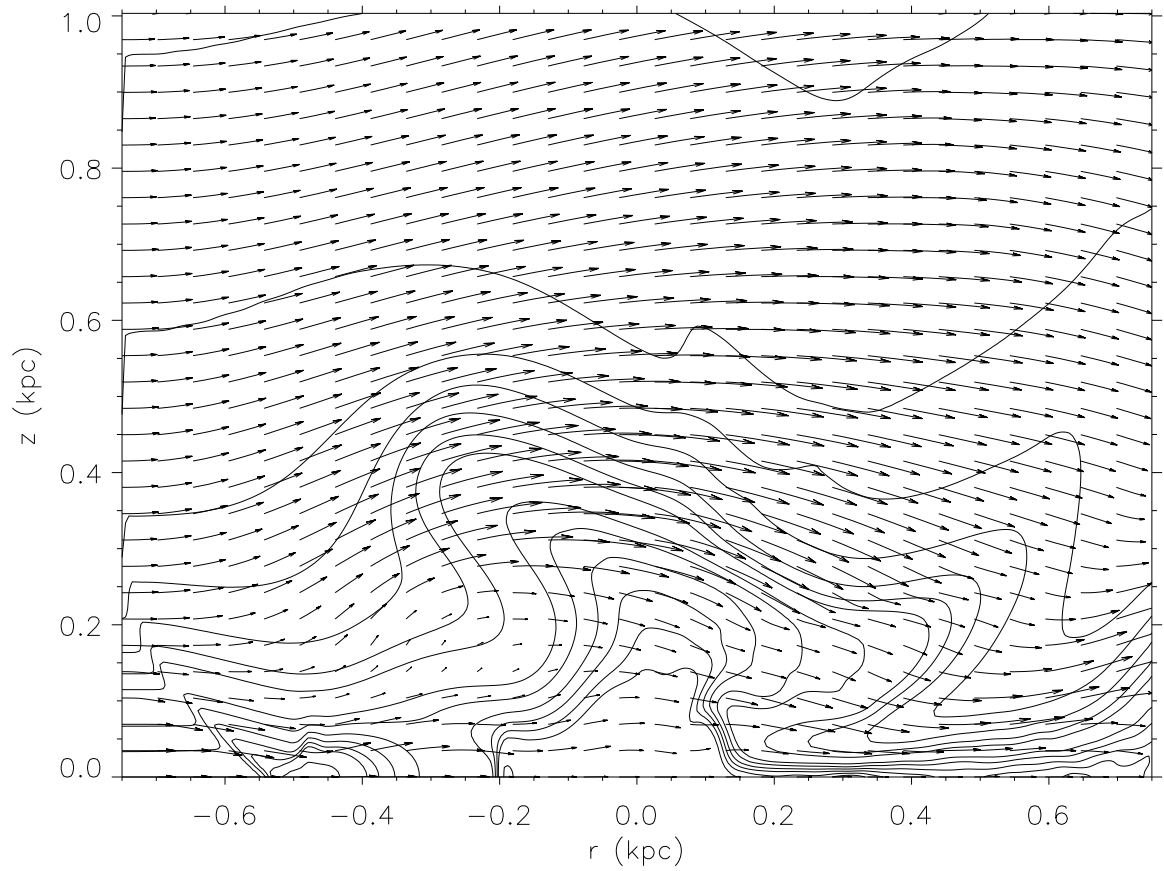
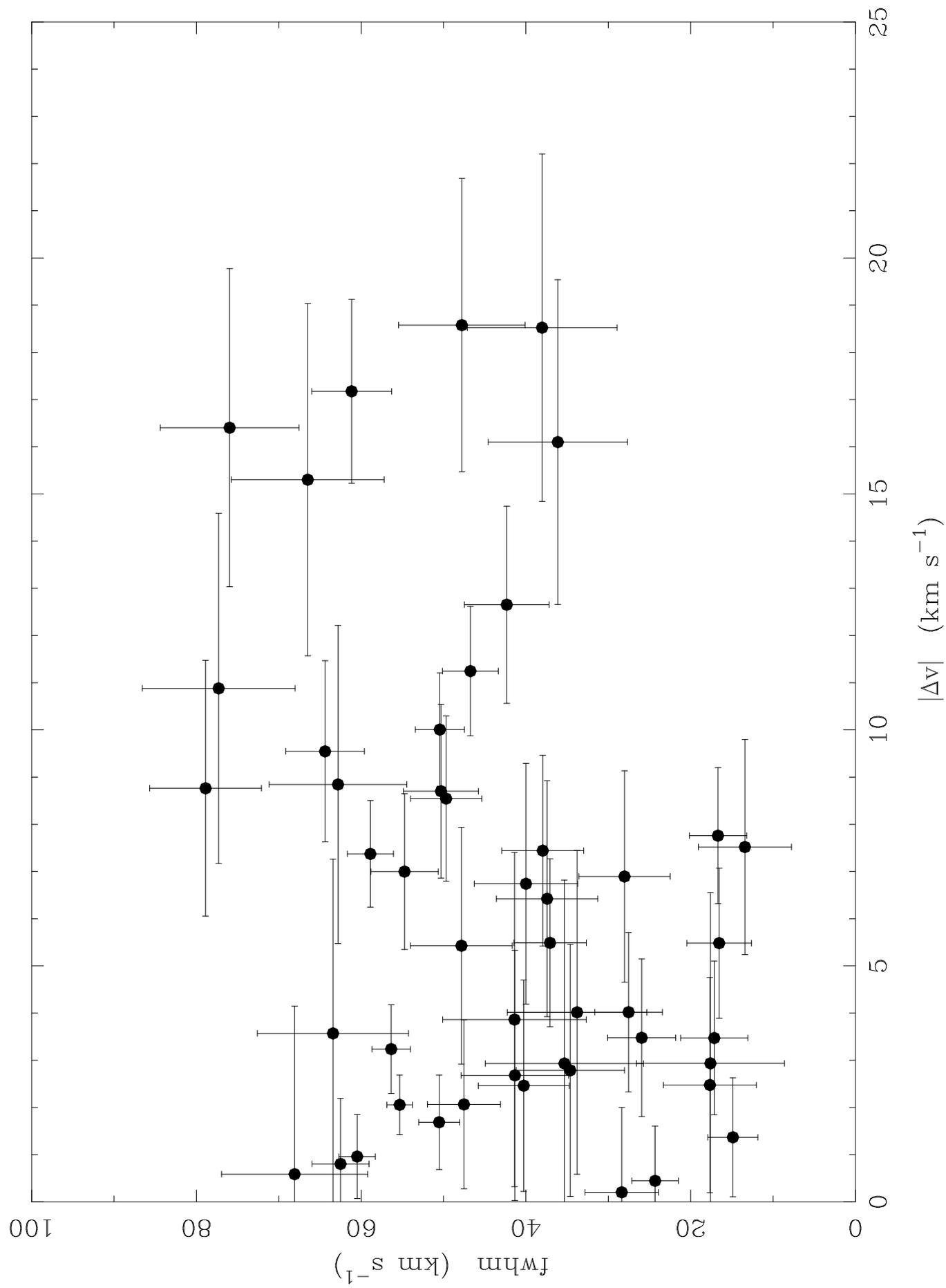
\end{document}